\documentclass{article}
\usepackage{frascatiphys_SP}
\newcommand{\be}{\begin{equation}}
\newcommand{\ee}{\end{equation}}
\newcommand{\beqy}{\begin{eqnarray}}
\newcommand{\eeqy}{\end{eqnarray}}

\newcommand{\etac}{\eta_c}
\newcommand{\as}{\alpha_s}
\newcommand{\etab}{\eta_b}

\begin{document}
\title{ TWO PHOTON WIDTH OF HEAVY PSEUDOSCALAR MESONS}
\author{Nicola~Fabiano\\
{\em Perugia University and INFN, via A.~Pascoli, I-06100 Perugia, Italy}\\
Giulia~Pancheri   \\
{\em INFN National Laboratories, P.O.Box 13, I00044 Frascati, Italy} }
\maketitle
\baselineskip=11.6pt
\baselineskip=14pt
\section{Introduction}
We discuss  the  partial width of the pseudoscalar charmonium 
state $\eta_c$ and bottomonium state $\eta_b$ into two photons.
Predictions from potential models are examined and 
compared with experimental values for the $\eta_c$ case.   Through the NRQCD
factorisation procedure results for $\eta_c$ are also compared with those 
from $J/\psi$ data, and results for $\eta_b$ to the $\Upsilon$ decay data.

\section{Experimental values and relation to vector electromagnetic width}
In this work we revisit the calculation of the two photon width of  $\etac$, 
highlighting newest experimental results and updating the potential model 
calculation. Unlike the $\etac$, the $\etab$ state hasn't been observed yet.
We will examine various theoretical predictions for the electromagnetic 
decay of the lightest $b \overline{b}$ bound state.
We start with the two photon decay width of a pseudoscalar 
quark--antiquark bound state with  first order QCD corrections.
The ratio of the pseudoscalar decay rate to the vector one is given by
\be
\frac{\Gamma (P\rightarrow \gamma\gamma)}{\Gamma(V
\rightarrow e^+e^-)}\approx
3Q^2 \frac{(1-3.38\as/\pi)}{(1-5.34\as/\pi)} = 
3Q^2 \left [ 1+1.96  \frac{\alpha_s}{\pi} 
+ \mathcal{O}(\alpha_s^2) \right ] \; .
\label{eq:rappwid}
\ee
This expression can be used  to estimate the radiative width 
of pseudoscalar state from the measured values of the leptonic decay width 
of the vector state. 

\section{Potential models predictions for $\eta_c$ and $\eta_b$ 
$\gamma \gamma$ decay width}
We present  the results one can obtain for the absolute width, through 
the extraction of the wave function at the origin from potential models.
The ``prototype'' potential is given by the Cornell potential model
\be
V(r) = -\frac{k}{r} + \frac{r}{a^{2}} 
\ee
which allows us to compute the Born decay width. The full expression of the
pseudoscalar decay width is given by:
\be
\Gamma(P \to \gamma \gamma) = \Gamma^P_{Born}
\left [ 1 + \frac{\alpha_s}{\pi}  \left(\frac{\pi^2-20}{3} \right ) \right ]
\ee
$\Gamma^P_{Born}$ depends on $\psi(0)$ of the particular potential model taken
into account. 
\section{Octet component procedure}
We  present after another procedure which admits other components to the 
meson  decay beyond the one from the colour singlet picture (Bodwin, Braaten 
and Lepage). 
The decay width expression is given by means of NRQCD from the expansion
\be
\Gamma = \sum_{n=1}^2 \frac{2\Im f_n(\as)}{M^{d{_n}}} \langle \mathcal{O}_n \rangle
\ee
which is a sum of terms, each of which factors into a short--distance
perturbative coefficient $\Im f_n$ and a long--distance nonperturbative
matrix element $\langle \mathcal{O}_n \rangle$. 
From the experimental values of electromagnetic decay and light hadron of the
vector state we obtain the two photon decay width of the pseudoscalar state.
We compute also the decay width from the lattice calculation of the
nonperturbative long--distance terms.

\section{Conclusions}
The $\Gamma(\eta_c \to \gamma \gamma)$ prediction gives the value $7.5 \pm 1.6
$ keV. The $\Gamma(\eta_b \to \gamma \gamma)$ prediction gives the value 
$466 \pm 101$ eV. Prediction of the BBL procedure and other theoretical 
results are in good agreement with each other.

\end{document}